\documentclass[10pt]{article}
\usepackage[table]{xcolor} 
\definecolor{lightgray}{gray}{0.9}
\usepackage[utf8]{inputenc}
\usepackage{fancyhdr}
\usepackage{extramarks}
\usepackage{amsmath}
\usepackage{amsthm}
\usepackage{siunitx}
\usepackage{tikz}
\usepackage[plain]{algorithm}
\usepackage{algpseudocode}
\usepackage{multirow}
\usepackage{booktabs}
\usepackage{graphicx}
\usepackage{subfigure}
\usepackage[colorlinks,linkcolor=black,anchorcolor=black,citecolor=black,urlcolor=blue]{hyperref}
\usepackage{amsmath,bm}
\usepackage{booktabs}
\usepackage{mathtools}
\usepackage{amssymb}
\usepackage{caption}
\usepackage{capt-of}
\usepackage{mciteplus}
\usepackage{cite}
\usepackage{mathrsfs}
\usepackage[title,titletoc,toc]{appendix}
\usepackage{xr}
\usepackage{parskip}
\usepackage{soul}
\usepackage{textcomp}
\usepackage[colaction]{multicol}
\usepackage[switch]{lineno}
\usepackage{lipsum}
\usepackage{etoolbox}
\usepackage{longtable}
\usepackage{array}
\usepackage{tablefootnote}
\usepackage{ragged2e}

\newcolumntype{C}[1]{>{\centering\arraybackslash}p{#1}}
\usetikzlibrary{automata,positioning}
\DeclareUnicodeCharacter{0308}{\"}
\title{PLD-Tree: Persistent Laplacian Decision Tree for Protein-Protein Binding Free Energy Prediction}

\author
{Xingjian Xu$^{1}$, Jiahui Chen$^{2,\dag}$  and Chunmei Wang$^{1, \ast}$\\
\normalsize{$^{1}$ Department of Mathematics,University of Florida, 1400 Stadium Rd, Gainesville, FL, 32611, USA}\\
\normalsize{$^{2}$Department of Mathematical Sciences, University of Arkansas, Fayetteville, AR 72701, USA}\\
\normalsize{$^\dag$ Address correspondences to Jiahui Chen. E-mail: jiahuic@uark.edu} \\
\normalsize{$^\ast$ Address correspondences to Chunmei Wang. E-mail: chunmei.wang@ufl.edu}\\
}

\date{}

\begin{document}


\baselineskip12pt


\maketitle

\begin{abstract}
Recent advances in topology-based modeling have accelerated progress in physical modeling and molecular studies, including applications to protein–ligand binding affinity. In this work, we introduce the Persistent Laplacian Decision Tree (PLD-Tree), a novel method designed to address the challenging task of predicting protein–protein interaction (PPI) affinities. PLD-Tree focuses on protein chains at binding interfaces and employs the persistent Laplacian to capture topological invariants reflecting critical inter-protein interactions. These topological descriptors, derived from persistent homology, are further enhanced by incorporating evolutionary scale modeling (ESM) from a large language model  to integrate sequence-based information.
We validate PLD-Tree on two benchmark datasets—PDBbind V2020 and SKEMPI v2 demonstrating a correlation coefficient ($R_p$) of 0.83 under the sophisticated leave-out-protein-out cross-validation. Notably, our approach outperforms all reported state-of-the-art methods on these datasets. These results underscore the power of integrating machine learning techniques with topology-based descriptors for molecular docking and virtual screening, providing a robust and accurate framework for predicting protein–protein binding affinities.
\end{abstract}

\section{Introduction}

Protein-protein interactions (PPIs) play a pivotal role in a wide array of biological functions in the human body, including cell metabolism, signal transduction, muscle contraction, and immune responses \cite{lu2020recent,mabonga2019protein,blazer2009small}. These complex interactions orchestrate cellular events that maintain homeostasis and respond to environmental changes. In therapeutic contexts, optimized PPIs are vital for the strong binding of antibodies to their protein antigens \cite{al2020antibody,bostrom2009improving,ausserwoger2022non,petta2016modulation,ben2007cancer}. This binding is fundamental to the efficacy of antibody-based drugs, which rely on precise molecular recognition to effectively target disease-causing agents. Precision medicine aims to tailor healthcare interventions to the individual characteristics of each patient, leveraging molecular insights to enhance therapeutic efficacy and minimize adverse effects, particularly by targeting specific protein-protein interactions crucial to disease mechanisms \cite{collins2015new, hood2012revolutionizing}. Therefore, characterizing PPIs in terms of their binding affinity is highly relevant to the design of new biologics and therapeutic compounds. By evaluating binding affinities, design therapies with improved specificity and reduced side effects, thereby offering innovative solutions for patients. 
However, accurately predicting the properties of PPIs remains challenging due to several factors, including the dynamic conformations of proteins, the impact of post-translational modifications, and the limitations of current computational models, which often struggle to fully capture the complexity of protein interactions and the diverse environments in which they function\cite{dill2012protein,jensen2004modification,feig2004mmtsb}. Furthermore, the vast structural diversity of proteins and variability in experimental conditions add layers of difficulty in developing reliable predictive models \cite{anfinsen1973principles,almo2013protein}.

The three-dimensional (3D) structure of PPIs provides essential topological and physical information, revealing binding sites, conformation changes, and the spatial arrangement of key residues critical for interactions. This structural insight forms the foundation for detecting biological properties and understanding underlying molecular mechanisms \cite{geng2019finding,goodsell2000structural}. 
The RCSB Protein Data Bank (PDB) is one of the largest and most comprehensive repositories of protein structures, including tens of thousands of protein-protein complex structures \cite{berman2000protein,rose2015rcsb}. As this resource continues to grow, it offers invaluable data that advance our understanding of the intricate interactions governing cellular processes.
Complementing structural data, thermodynamic measurements provide quantitative insights into interaction strength and stability, such as binding affinity, enthalpy, and entropy changes. By integrating structural and thermodynamic data, one can gain a comprehensive view of PPIs, facilitating the design of inhibitors, activators, and therapeutic agents. 
To this end, techniques such as X-ray crystallography, nuclear magnetic resonance (NMR) spectroscopy, and cryo-electron microscopy (cryo-EM) have been developed to determine the structures of protein-protein complexes, and these are among the primary methods employed in structural biology \cite{wuthrich2003nmr,cheng2018single}.


With the rapid development of techniques in structural biology and computational modeling, the availability and diversity of datasets for studying PPIs have expanded significantly, providing an ever-growing resource for advancing predictive and analytical methods. These datasets can be broadly categorized into two types based on their focus: binding free energy data ($\Delta G$) and mutation-induced binding free energy changes ($\Delta\Delta G$). The first type comprises data on the binding free energy of protein-protein interactions, which is crucial for understanding the inherent stability and affinity of interactions under normal physiological conditions. 
For example, the PDBbind database provides the binding affinities data for various types of protein complexes which contains { 2852 complexes in 2020 version }\cite{liu2015pdb,su2018comparative,liu2017forging}.
However, while PDBbind has been extensively utilized for studying protein-ligand binding affinities\cite{nguyen2019agl,volkov2022frustration,cang2018representability,ballester2010machine}, the subset focusing on protein-protein interactions remains underutilized due to the limited availability of binding partner information and incomplete dataset collections. This underutilization highlights the need for expanded and curated datasets to fully unlock the potential of predictive models for PPIs.
The latter focuses on mutation-induced binding free energy changes, providing measurements of binding free energy for both wild-type and mutant forms of interacting proteins. These datasets are particularly valuable for understanding how specific mutations can enhance or disrupt protein function and are widely used in protein engineering to design proteins with improved or altered functionalities\cite{sirin2016ab,moal2012skempi,jankauskaite2019skempi}. 
For example, SKEMPI is a manually curated data repository that catalogs {7,085 } mutations in experimentally determined structurally characterized protein-protein interactions and their effects on binding affinity \cite{moal2012skempi,jankauskaite2019skempi}. These datasets serve as essential benchmarks for evaluating the predictive power of computational methods. They are particularly indispensable for investigating PPIs, providing critical training sets for systematic screening of mutations in real-world applications, such as analyzing the impact of spike protein mutations in SARS-CoV-2.



Computational methods have become essential tools for calculating the binding affinity of PPIs, drawing upon the principles of molecular dynamics and quantum mechanics. Rigorous approaches such as free energy perturbation (FEP) and thermodynamic integration have been utilized to compute PPI interaction energies; however, these methods are computationally expensive and often face convergence challenges due to the necessity of simulating numerous nonphysical intermediate states \cite{bash1987free,jorgensen2008perspective,kita1994contribution,kollman1993free,zacharias1994separation}. To address these limitations, alternative techniques like linear interaction energy and molecular mechanics methods, notably the MM/GBSA approach, have gained popularity for their balance between accuracy and computational efficiency \cite{massova1999computational,huo2002computational,obiol2008protein,zuo2012free,ylilauri2013mmgbsa}. Additionally, empirical methods—including force-field potentials, statistical potentials, and scoring functions used in protein docking—have evolved to offer faster predictions and have paved the way for integrating machine learning techniques to enhance accuracy without significantly increasing computational costs
\cite{vangone2015contacts,ravikant2010pie,abbasi2020island,yugandhar2014protein,huang2020ssipe,abbasi2018learning,liu2004physical,wang2021computational,bryant2022improved,cheron2017update}. 
For instance, tools like PRODIGY leverage structural and functional features, such as networks of interfacial contacts, to predict binding affinities using regression models\cite{vangone2015contacts}. CP\_PIE employs scoring functions and includes benchmark datasets for evaluation, enhancing its utility in predicting PPI affinities\cite{ravikant2010pie}. Techniques like DFIRE utilize statistical potentials derived from docking studies to assess protein-peptide and protein-protein complexes\cite{liu2004physical}. Machine learning approaches have also been integrated, with methods like PPI-Affinity using support vector machines and MmCSM-PPI applying Monte Carlo simulations to reconstruct binding affinities by decomposing contributions from interfacial and non-interfacial residues\cite{wang2021computational}. Sequence-based methods, such as ISLAND\cite{abbasi2020island} and PPA-pred\cite{yugandhar2014protein}, predict binding affinities using amino acid sequence information, expanding the toolkit available for PPI analysis. A minimal yet crucial discussion of binding site identification methods underscores the importance of accurately determining interaction interfaces, which is fundamental for improving the reliability of binding affinity predictions.

Persistent homology, a cutting-edge area within algebraic topology, bridges geometry and topology to unveil the underlying structures of complex systems\cite{edelsbrunner2010computational,zomorodian2004computing}. Element-specific persistent homology addresses limitations by preserving essential biological details during topological abstraction, thereby facilitating a deeper understanding of protein-protein interactions upon mutations \cite{wang2020topology}. This approach offers innovative ways to characterize and comprehend the complex topological features of protein complexes. However, persistent homology alone may not be sufficient for representing protein complex data comprehensively. This challenge in TDA was addressed by the introduction of the persistent Laplacian, or persistent spectral graph theory \cite{wang2020persistent}. The persistent Laplacian manifests the full set of topological invariants and captures the shape of data through its harmonic and non-harmonic spectra, respectively. Additional mathematical analyses \cite{memoli2020persistent} and software packages like HERMES \cite{wang2021hermes} have been developed to support persistent Laplacian methods. This approach has been successfully applied to biological studies, including protein thermal stability \cite{wang2020persistent}, protein-ligand binding \cite{meng2021persistent}, and protein-protein binding with mutation problems \cite{chen2022persistent}.

This study introduces PLD-Tree, a novel approach that integrates topology-based methods with physical attributes ,which can indeed represent the efficient information of
protein-protein complexes. Specifically, by focusing on specific binding interfaces across
various complex types, we construct persistent homology for $H_0$, $H_1$, and $H_2$ groups using atomic and amino acid representations and utilize ESMFold structural predictions \cite{rives2019biological} and physical interaction forces to extract and characterize additional efficient features  to predict the binding affinities of PPIs. The PLD-Tree model is evaluated on two benchmark molecular datasets, and extensive validations and comparisons demonstrate that it yields some of the most accurate predictions of PPI properties.

\section{PLD-Tree Model for Prediction}
\begin{figure}[ht]
    \centering
    \includegraphics[width=1\textwidth]{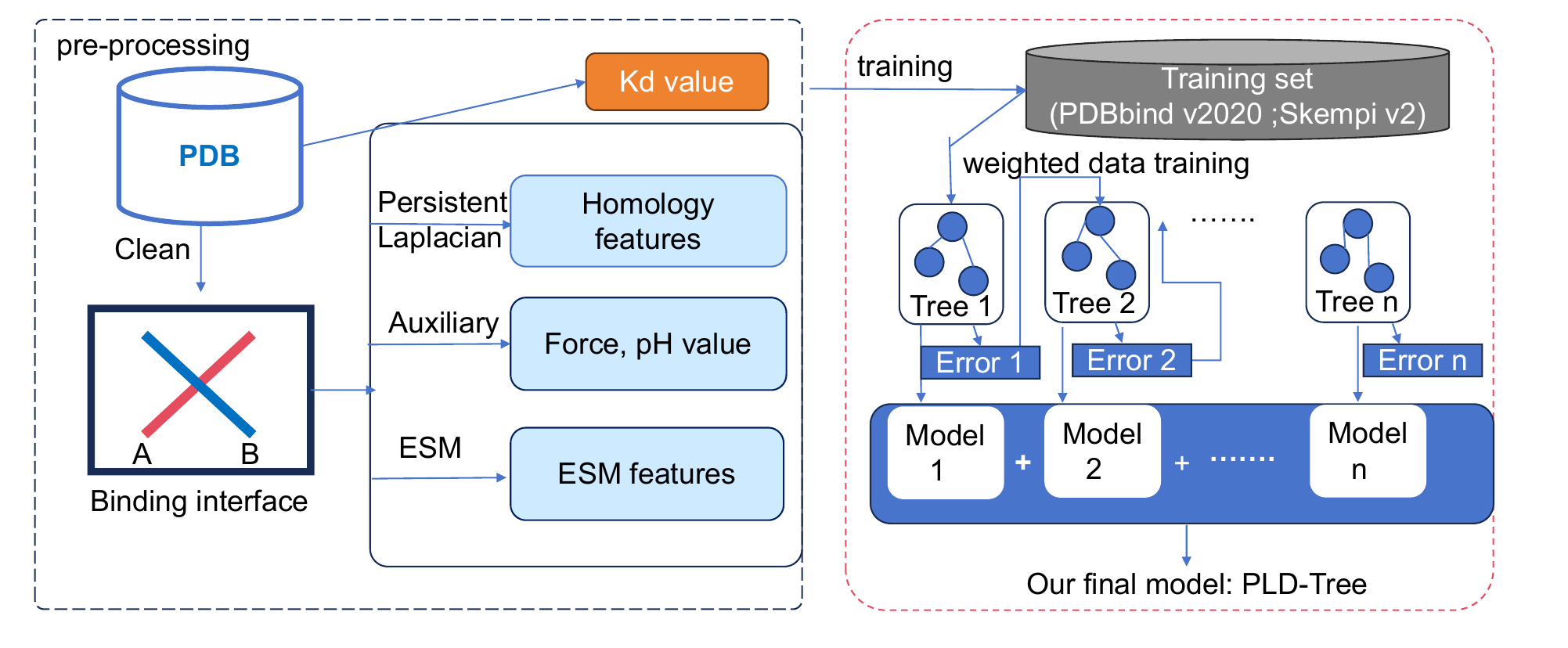}
    \caption{Structure of PLD-Tree model. Protein structures are first preprocessed to define binding interfaces (A and B) and retrieve their measured Kd values. From these interfaces, three feature sets are extracted: persistent Laplacian (homology) features, auxiliary (force, pH) features, and ESM embeddings. These features then serve as input to a gradient-boosting ensemble of decision trees, culminating in the final PLD-Tree predictor for protein–protein binding affinities.}
    \label{fig:structure_paper}
\end{figure}

This section outlines the PLD-Tree model and its algorithm for predicting protein-protein interaction (PPI) binding free energy ($\Delta G$). As depicted in Figure \ref{fig:structure_paper}, the model consists of two primary modules: topology-based feature generation and a gradient boosting decision tree (GBDT) model. The feature generation module leverages element- and residue-specific topological features augmented with chemical-physical descriptors to capture structural characteristics effectively. The performance of the PLD-Tree model is validated on two benchmark datasets—PDBbind V2020 and SKEMPI v2. Figure \ref{fig:structure_paper} further illustrates how pairwise interactions between atoms are characterized using the zeroth homology group ($H_0$) for clusters and the first homology group ($H_1$) for loop-like structures derived from Euclidean distance-based filtration.

\subsection{Data Collection: PDBbind V2020 and SKEMPI}
We employed two primary datasets, PDBbind V2020 and SKEMPI v2, as the foundation for training and evaluation. The PDBbind V2020 dataset originally contained 2,852 protein-protein complexes with experimentally measured binding affinities. To ensure consistency, we retained only complexes with affinities reported as $K_d$, $K_i$, or $\Delta G_{\text{bind}}$. Complexes with incomplete physical property data or imprecise binding values (e.g., reported as a range rather than a specific number) were either excluded or approximated using boundary estimates. Additionally, because PDBbind v2020 lacks explicit binding partner information, we verified each PPI complex individually through the RCSB PDB database to identify the correct partners. After these steps, a total of 2,571 complexes remained. Each complex underwent structural preprocessing with Profix\cite{xiang2001extending}, which restored missing residues and improved data completeness. For each refined PPI complex, we identified the binding interface and designated the interacting partners as “Partner A” and “Partner B,” a classification that was later leveraged to enhance model accuracy.

The SKEMPI v2 dataset, an expanded version of its predecessor, integrates new mutations from AB-Bind, PROXiMATE, and dbMPIKT databases. In total, it includes 7,085 mutation-induced binding affinity changes ($\Delta\Delta G$) across 345 protein-protein interactions with resolved complex structures. Following a similar curation process—excluding ambiguous mutants and using arithmetic means for repeated measurements—we arrived at a set comprising 342 wild-type systems and 4,175 single mutants. SKEMPI v2 inherently classifies binding partners for each interaction, thus eliminating the need for further subdivision into partner A and B. Additionally, the mutant structures were generated by introducing a single amino acid substitution into the corresponding wild-type PDB structure.

{\bf Representation of Binding Interfaces.}
After data refinement, we focused on effectively representing the binding interfaces present in these curated complexes. Each PPI complex was considered as two interacting sets, $A$ and $B$, corresponding to the designated Partner A and Partner B identified during data processing. We targeted all atoms within a defined cutoff distance $r$ surrounding the interface, capturing the critical region of intermolecular contact. Within these selected atoms, two classifications can be used to facilitate analysis:
\begin{enumerate}
    \item Element-specific sets: Groups of atoms categorized by their elemental composition.
    \item Residue-specific sets: Groups of atoms organized according to the amino acid residues they belong to.
\end{enumerate}

{\bf Modeling Non-Covalent Interactions.}
To systematically analyze and quantify the non-covalent interactions at the binding interface, we employed a modified distance function $D_{\text{mod}}(a_i, a_j)$:
\begin{equation}
D_{\text{mod}}(a_i, a_j) =
\begin{cases}
\infty, & \text{if } a_i, a_j \in A \text{ or } a_i, a_j \in B \\
\|\mathbf{r}_i - \mathbf{r}_j\|, & \text{if } a_i \in A \text{ and } a_j \in B
\end{cases}
\label{modifiedditance}
\end{equation}
where, $\mathbf{r}_i$ and $\mathbf{r}_j$ are the position vectors of atoms $a_i$ and $a_j$, respectively, and $|\mathbf{r}_i - \mathbf{r}_j|$ is the Euclidean distance between them, which we denote as $\Phi_{ij}$ for convenience.
In this framework, “Partner A” and “Partner B” represent the primary components of the PPI complex. When multiple chains are present in a single protein, the following scenarios clarify how we define these sets:
\begin{enumerate}
    \item If only one chain in a multichain protein participates in binding, we treat it as both Partner A and Partner B, effectively representing a self-binding scenario.
    \item If multiple chains are present but do not actually form a binding interface, such a non-binding scenario is not addressed.
    \item If multiple chains interact to form more than one binding interface, we focus only on the primary or most inclusive binding interface and do not consider additional minor interfaces.
\end{enumerate}
By applying the modified distance function $D_{\text{mod}}$ to these well-defined sets of atoms (either element-specific or residue-specific), we can systematically capture the key non-covalent interactions that govern protein-protein binding. This representation serves as a crucial step for downstream modeling tasks, including feature extraction and topological analyses aimed at predicting binding affinities.

\subsection{Topological Embedding Features}
Topological embedding features leverage the geometric and structural complexities of protein-protein interfaces by examining point clouds derived from atomic coordinates or amino acid positions. Through this lens, a wide array of topological invariants, including $H_0$ (connected components), $H_1$ (loops or cycles), and $H_2$ (voids or cavities), can be extracted. Persistent homology provides a powerful mechanism for capturing these invariants across varying scales, while the persistent Laplacian further refines the analysis by integrating geometric considerations through targeted filtration processes. This combined framework yields a rich topological characterization of binding interfaces, offering deeper insights into the underlying architecture of PPIs.

Using a Rips complex constructed from the distance matrix of these complexes, we derive persistence diagrams that highlight critical topological features—clusters, holes, and other structural motifs—across multiple scales. To hone in on the most meaningful invariants, we apply a cutoff and filter out barcodes with lifespans, ensuring that only the most significant topological patterns are retained. Subsequently, barcodes falling within specified death ranges are cataloged and binned, enhancing our capacity to elucidate the topological landscape of the binding interface.
To further dissect these topological features, we examine subsets of the distance matrix at progressively tighter distance thresholds (Bins), generating Laplacian matrices whose eigenvalues encode geometric and topological information. For each pair of elements in a given element list, we compute statistical descriptors of these eigenvalues, including their sum, minimum, maximum, mean, standard deviation, variance, sum of squares, and the count of significant eigenvalues. Such a statistical treatment of eigenvalues provides a multifaceted characterization of the complexes’ structural and topological signatures.

This comprehensive methodological approach, outlined in Eq. \eqref{modifiedditance}, enables the systematic examination of persistent homology and topological features at the atomic level. Expanding the same principles to the amino acid level supplies yet another layer of detail, capturing subtle differences in topological features arising from residue-level organization. Ultimately, this topological embedding framework allows us to identify and quantify robust patterns that define the binding interfaces of PPIs, paving the way for more accurate predictions of binding free energies and an improved understanding of protein function.

\subsection{ESM Transformer Features}
Recent advancements in protein property modeling have emerged from the use of large language models trained on extensive protein sequence datasets. Models such as Evolutionary Scale Modeling(ESM)\cite{lin2022language,hsu2022learning,meier2021language} and ProtTrans\cite{9477085} have exhibited remarkable capabilities, which can be further enhanced by hybrid fine-tuning strategies that incorporate both local and global evolutionary signals. In particular, the ESM model can be fine-tuned using downstream task data or local multiple sequence alignments, yielding increasingly accurate predictions. In this study, we utilized the ESM-2 (t33 650M UR50D) transformer model, originally trained on a dataset of 250 million protein sequences using a masked sequence prediction objective. With a 34-layer architecture and approximately 650 million parameters, ESM-1b generates rich sequence embeddings. For sequences that exceed the model’s input length, we partitioned them into subsequences, ensuring that accurate embeddings could still be derived. This approach produced a 3,840-dimensional feature vector for each PPI complex, substantially improving the performance and predictive power of our model.

\subsection{Auxiliary Features}
In addition to sequence-derived features, we also integrated a suite of biochemical and biophysical descriptors to capture the molecular interactions governing PPIs. These features included solvent-accessible surface area (ASA), which quantifies the portion of the protein’s surface in contact with the solvent, and buried surface area, which identifies regions concealed upon complex formation. Assigning partial charges to interacting residues using methods like PDB2PQR\cite{dolinsky2004pdb2pqr} enabled us to compute Coulombic interaction energies via Coulomb’s law. The Lennard-Jones potential further refined our representation by accounting for van der Waals interactions, balancing attractive and repulsive atomic forces.
Additionally, we solved the Poisson-Boltzmann equation to incorporate electrostatic effects, reflecting a physiological ionic environment and fixed protein charges using MIBPB \cite{chen2011mibpb}. This calculation yielded electrostatic potential maps and electrostatic free energies of binding, which served as another valuable set of features. In total, our model extracted 74 such biophysical features for Partner A and Partner B, respectively. By integrating these comprehensive physical descriptors with sequence-derived embeddings, we constructed a more robust, informative representation of PPIs, ultimately leading to deeper insights and improved predictive accuracy. Details of auxiliary feature generation are presented in the supporting information.





\subsection{Machine Learning Models}
Our predictive modeling approach utilized Gradient Boosting Decision Trees (GBDT), a proven and versatile machine learning algorithm widely employed in drug discovery and other domains. We initially explored various models, including Multilayer Perceptrons (MLPs), but GBDT consistently outperformed them in terms of predictive accuracy and stability. The superior performance of GBDT, combined with its robustness to overfitting, made it our method of choice for Kd value prediction of PPI complexes.
GBDT builds an ensemble of weak learners (decision trees) in a sequential manner, with each tree reducing the residual errors of the previous ones. This approach often achieves performance comparable to that of random forests, while also providing flexibility through a rich set of tunable hyperparameters. To optimize GBDT performance, we conducted a grid search over key hyperparameters { of sklearn \cite{scikit-learn}}. The optimal configuration was:
\texttt{n\_estimators = 25000, max\_depth = 7, min\_samples\_split = 3, learning\_rate = 0.001, subsample = 0.3, and max\_features = sqrt}.
Small variations around these parameters did not significantly impact predictive accuracy, indicating a stable and well-tuned model.




\section{Results and Discussion}
In the following section, we evaluate our models using the PDBbind V2020 dataset as the foundational training set. We then augment this dataset by incorporating wild-type (wt) and mutant (mt) complexes from SKEMPI v2 in three different scenarios. These augmented datasets enable a broader performance comparison against other state-of-the-art predictors on their respective benchmark sets. When evaluating models on mutation-based datasets, multiple mutation complexes may originate from the same protein complex. To ensure independence and fairness in our performance assessment, we employ a leave-one-protein-out validation strategy. This approach prevents information leakage between training and testing subsets that would otherwise skew results.

Our evaluation metrics include Pearson’s correlation coefficient ($R_p$) and Mean Absolute Error (MAE). We compute $R_p$ not only on the training data (10-fold cross-validation) but also on an independent development set. By combining both correlation and error metrics, we ensure a comprehensive understanding of the model’s predictive capabilities. Pearson’s correlation coefficient $R_p$ is defined as:
\begin{equation}
R_p = \frac{\sum_{i=1}^{n}(y_i-\bar{y})(y_{i}^{\text{pred}}-\bar{y}^{\text{pred}})}{\sqrt{\sum_{i=1}^{n}(y_i-\bar{y})^2 \sum_{i=1}^{n}(y_{i}^{\text{pred}}-\bar{y}^{\text{pred}})^2}}
\end{equation}
Here, $y_i$ and $\bar{y}$ represent the actual affinity values and their mean, respectively, while $y_{i}^{\text{pred}}$ and $\bar{y}^{\text{pred}}$ are the predicted values and their mean. By using both correlation and error-based metrics, we gain a thorough, reliable perspective on model performance, ultimately ensuring meaningful predictions of binding affinities.



\subsection{Performance of the Protein–Protein Benchmark Model}
In this section, we present additional experiments that illustrate the model’s versatility and further validate its performance.
We conducted three benchmark tests to validate our proposed PLD-Tree model for utilizing three distinct benchmark datasets, designated as S79\cite{vangone2015contacts}, S90\cite{vangone2015contacts, romero2022ppi}, and S177\cite{romero2022ppi}. The first benchmark set, S79, is from the web server PRODIGY \cite{vangone2015contacts}. The second benchmark set, S90, is a carefully curated subset of the refined PDBbind V2020 dataset \cite{romero2022ppi}. The final benchmark set, S177, encompasses 26 wild-type PPI complexes alongside 151 mutant complexes\cite{romero2022ppi}. By evaluating PLD-Tree across these three benchmark datasets, we ensure a comprehensive validation of its predictive performance in diverse and challenging contexts. This multi-dataset approach allows us to compare our model against existing state-of-the-art methods effectively, demonstrating its robustness and versatility in handling both wild-type and mutant PPI complexes. 

\begin{table}[ht]
\centering 
\caption{The performance for $R_p$ and MAE between experimental and predicted binding affinities on the S79 and on the S90.}
\label{tab:comparison}
\begin{tabular}{|l|c|c|c|c|}
\hline
\rowcolor{lightgray} & \multicolumn{2}{c|}{S79} & \multicolumn{2}{c|}{S90} \\  \hline
\rowcolor{lightgray} {Method} & \textbf{$R_p$} & MAE (kcal/mol)& \textbf{$R_p$} & MAE (kcal/mol)\\  \hline
PRODIGY \cite{xue2016prodigy}& 0.74& 1.4 & 0.31 & 2.5\\ \hline
DFIRE  \cite{liu2004physical}  & 0.60 & 4.6 & 0.10 & 25.4 \\ \hline
CP$\_$PIE \cite{ravikant2010pie}    & -0.50 & 8.8 & -0.10 & 11.0\\ \hline
ISLAND \cite{abbasi2020island}     & 0.38 & 2.1 & 0.27 & 2.2\\ \hline
PPI-Affinity \cite{romero2022ppi} & 0.62 & 1.8 & 0.50 & 1.8\\ \hline
\textbf{PLD-Tree} & {\textbf{0.68}} & {\textbf{1.4}} & \textbf{0.70} & \textbf{1.4}\\ \hline
\end{tabular}
\footnotesize
\end{table}
The ICs/NIS-based predictor is implemented in the web server PRODIGY \cite{vangone2015contacts}, which estimates binding affinity (BA) using two structural descriptors: the network of inter-residue contacts (ICs) and the noninteracting surface (NIS). On a benchmark set of 79 protein–protein complexes, PRODIGY achieved a $R_p$ of 0.74 and an MAE of 1.4 kcal/mol, outperforming other state-of-the-art methods. To evaluate the performance of our PLD-Tree model, we employed the same benchmark set (Table \ref{tab:comparison} test set S79) and compared it against PRODIGY, three other top-ranked tools available at the time, and the ISLAND method. Our ensemble PLD-Tree model attained a Pearson correlation coefficient of $R_p = 0.68$ and an MAE of 1.4 kcal/mol on the S79 test set. This performance ranks our method second, closely trailing PRODIGY, while maintaining an identical MAE. Although our model exhibits a slightly lower correlation with experimental data compared to PRODIGY, the equivalent MAE indicates that PLD-Tree achieves comparable accuracy in predicting binding affinities. This demonstrates the effectiveness of our approach, particularly in maintaining prediction precision despite the lower correlation. 

Furthermore, we benchmarked our PLD-Tree model against additional test sets from \cite{romero2022ppi} to comprehensively evaluate its performance. Test Set S90, a subset of the refined PDBbind V2020 and SKEMPI v2 datasets already utilized in our study, demonstrated notably high $R_p$ for our model. As shown in Table \ref{tab:comparison}, when trained using the same dataset as \cite{romero2022ppi}, our PLD-Tree model achieved a performance that significantly surpasses existing methods, particularly non-machine learning approaches ($R_p=0.57$, MAE$=1.6$kcal/mol). Specifically, PLD-Tree attained an $R_p=0.70$ and an MAE of $1.6$ kcal/mol on S90. Despite this slight decrease, our model still outperformed other predictors, with ISLAND showing diminished performance relative to PLD-Tree. 
In contrast, other predictors such as PRODIGY, DFIRE, and CP\_PIE experienced a substantial decline in their performance on Test Set S90 compared to Test Set S79 (Table \ref{tab:comparison}). 
For instance, PRODIGY's performance dropped to $R_p = 0.31$ and MAE $= 2.5$ kcal/mol, highlighting potential overfitting issues towards the original benchmark set. 
This dramatic decay suggests that these models may not generalize well to more diverse or larger datasets. Conversely, PPI-Affinity maintained relatively consistent performance across both test sets, although it still did not match the PLD-Tree’s results.

The superior performance of PLD-Tree on Test Set S90 underscores the advanced predictive capabilities of our persistent homology-based machine learning approach for PPIs. Moreover, our model’s $R_p$ exceeding 0.50 not only outperforms the previously best-known result reported by PPI-affinities \cite{romero2022ppi} but also reaffirms the effectiveness of the PLD-Tree method in accurately predicting binding affinities. However, the analysis of other predictors is complicated by the absence of a defined applicability domain, which limits our ability to determine whether test samples fall outside the models’ effective scope or if specific structural factors adversely impact prediction quality. Nonetheless, the robust performance of PLD-Tree across multiple benchmark sets highlights its potential as a reliable and generalizable tool for PPI binding affinity prediction.

\begin{figure}
\centering
\includegraphics[width=0.4\linewidth]{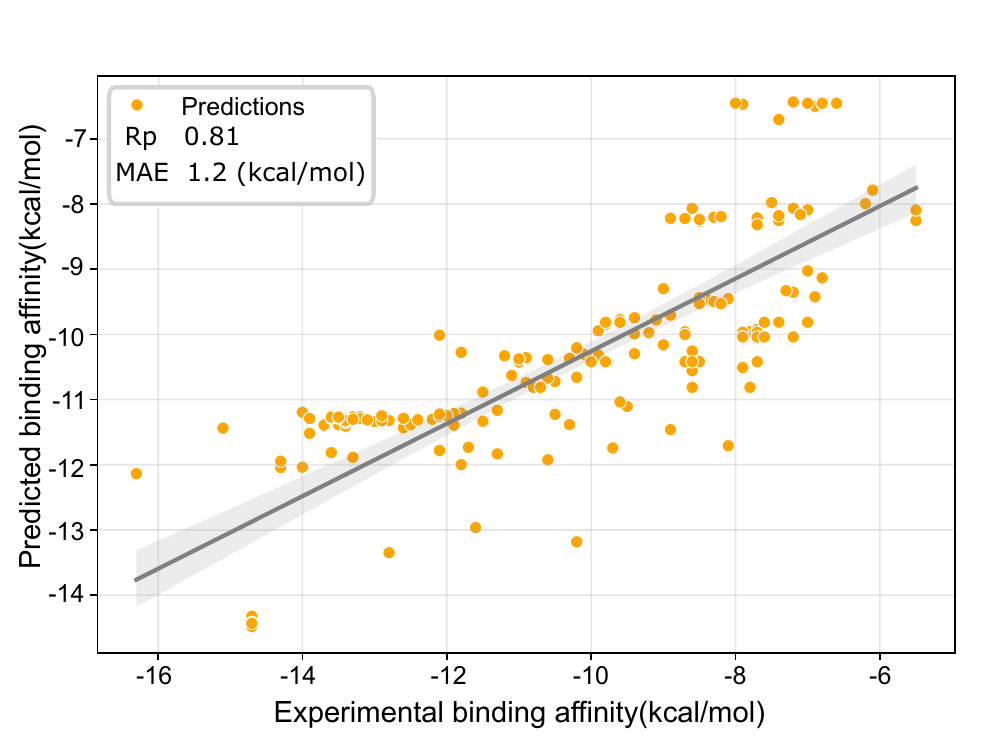}
\caption{The performance for $R_p$ of 0.81, MAE of 1.2 kcal/mol and scatter plot between between experimental and predicted binding affinities on S177}
\label{fig:test set 3}
\end{figure}    
The third benchmark set, S177, is derived from PPI-Affinity \cite{romero2022ppi}. The mutant complexes in this set feature between one and six mutations per protein sequence, with approximately 80\% of the structures containing only a single mutation. The binding free energies of all complexes within S177 range from -16.3 to -5.5 kcal/mol, providing a diverse spectrum of affinity values for assessing the model’s accuracy across various mutation scenarios. Our PLD-Tree model achieves a $R_p$ of 0.81 and an MAE of 1.24 kcal/mol (Figure \ref{fig:test set 3}), compared to PPI-Affinity’s $R_p = 0.78$ and MAE $= 1.40$ kcal/mol. Note the PLD-Tree model has a better performance on other training datasets ($R_p = 0.85$), while the reported dataset has achieved a high correlation in Figure \ref{fig:test set 3}. These results demonstrate that our approach surpasses all benchmark methods presented in the referenced study. Notably, even without explicit training on data containing multi-mutant complexes, our model delivered compelling and superior predictive performance. This further highlights the robustness and versatility of our PLD-Tree framework in handling diverse and complex PPI scenarios. The robust performance of our model underscores its potential as a competitive alternative to existing methods, emphasizing the value of our feature optimization and integration strategies in enhancing predictive capabilities for protein–protein interactions.

\subsection{Performance of Training Sets}   
\begin{figure}
\begin{center}
\includegraphics[width=0.6\linewidth]{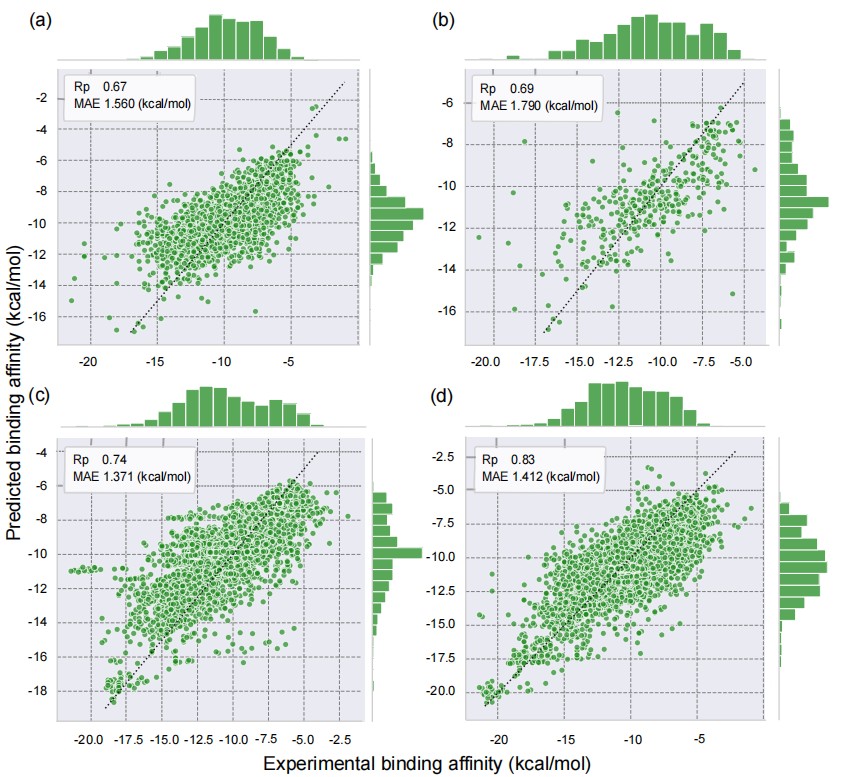}
\end{center}
\caption{The performance of scatter plots using PLD-Tree model on the training sets for PDBbind V2020, and SKEMPI v2.}
\label{fig:P2P}

\end{figure}

We evaluated PLD-Tree's performance using 10-fold cross-validation on the considered datasets, as shown in Table \ref{tab:comparison 1} and Figure \ref{fig:P2P}. Notably, we applied a leave-one-protein-out strategy to all SKEMPI v2 subsets—whether wild-type (wt), mutant (mt), or combined—since each protein in SKEMPI v2 can harbor multiple mutations.  For cross-validation, we therefore grouped entire proteins into the same fold to prevent data leakage between training and testing. When trained and tested solely on the PDBbind V2020 dataset, PLD-Tree achieved a Pearson correlation coefficient ($R_{p}$) of 0.672 and a mean absolute error (MAE) of 1.560 kcal/mol, providing a solid baseline for subsequent enhancements. By augmenting the dataset with wt and mt complexes from SKEMPI v2, we observed a considerable increase in predictive performance, with PLD-Tree reaching an $R_{p}$ of 0.83 and an MAE of 1.412 kcal/mol on the combined PDBbind V2020–SKEMPI v2 dataset.

Moreover, merging the wt and mt data yielded an even higher cross-validation performance ($R_{p} = 0.87$ and MAE = 1.115 kcal/mol), despite maintaining the leave-one-protein-out design. Figure \ref{fig:P2P} presents scatter plots for the three individual sets and their combined dataset, illustrating consistently strong results across each partition. Although the mt dataset exhibits the highest correlation (Figure \ref{fig:P2P}(c)), a few outliers remain visible; by contrast, the combined dataset demonstrates fewer outliers and further improves cross-validation accuracy. This substantial gain in predictive power surpasses previously reported results and underscores the effectiveness of our feature optimization strategy. To our knowledge, this is the first time such a high correlation has been achieved on the PDBbind V2020 dataset, highlighting PLD-Tree’s robust ability to integrate diverse data sources and more accurately capture the underlying determinants of protein-protein binding affinity.


\begin{table}[ht]
\centering 
\caption{The performance for $R_p$ and MAE between experimental and predicted binding affinities on training sets for PDBbind V2020, SKEMPI v2, and SKEMPI v2 wild type (wt) and mutant (mt) are presented separately. }
\label{tab:comparison 1}
\begin{tabular}{|l|c|c|}
\hline
\rowcolor{lightgray} Datasets & \textbf{$R_p$} & MAE(kcal/mol)  
\\  \hline
PDBbind V2020 &  

{{0.67}} &  {{1.560}} \\ \hline

{SKEMPI v2 wt} & {0.69}& {{1.790}} \\ \hline

{SKEMPI v2 mt} & {{0.74}} & {{1.371}} \\ \hline

{SKEMPI v2}  & {{0.87}} & {{1.115}} \\ \hline


{{PDBbind V2020-SKEMPI v2}} & 

{{0.83}} & {{1.412}}\\ 
\hline
\end{tabular}
\footnotesize
\end{table}

\subsection{Performance of Binding Partners Classification for PDBbind V2020}
\begin{table}[ht]
    \centering
    \caption{The size of datasets on PDBbind V2020 grouped by different numbers of binding proteins. }
    \begin{tabular}{|c|c|c|c|c|c|c|}
    \hline
    \rowcolor{lightgray} Binding proteins & 1 & 2 & 3 & 4 & 5 & $>$5 \\
    \hline
  Dataset Size  & 6 & 2414 & 123 & 190 & 107 & 12 \\
    \hline
    \end{tabular}
    \label{tab:num_binding}
\end{table}
Unlike the SKEMPI dataset, the PDBbind V2020 dataset is more recent and does not provide explicit binding partner information. Before partitioning all PPI affinities at the binding interface into two partners—a process further complicated by symmetrical structures and multiple protein chains—we first relied on three-dimensional structural data from the RCSB PDB. We then categorized the PPI complexes in the V2020 dataset into different binding classes. As summarized in Table \ref{tab:num_binding}, the dominant category features two binding partners (2,414 complexes), many of which exhibit symmetrical structures involving DNA, RNA, or viral components.

\subsection{Performance for Antibody-Antigen Classification of PDBbind V2020}
Developing new pharmaceutical compounds is a lengthy, costly, and resource-intensive process. In support of drug design, it is also valuable to assess datasets consisting exclusively of antibody and antigen structures. Accordingly, we identified each chain in the PDBbind v2020 database and subset wt of SKEMPI v2 database to extract 630 relevant complexes (hereafter denoted S630), as well as an additional 55 wild-type records (hereafter S55). Detailed information about these datasets can be found in the Supplementary Information.

We then evaluated PLD-Tree’s performance using 10-fold cross-validation on these antibody–antigen datasets and the performance are detailed in the Supporting Information. 
When trained and tested solely on S630, PLD-Tree achieved a $R_p$ of 0.48 and an MAE of 1.322 kcal/mol. However, after augmenting S630 with S55 to form S685, we observed an improvement, reaching $R_p = 0.51$ and an MAE of 1.248 kcal/mol. While these results indicate potential for practical application, we anticipate that further refinements to feature selection and modeling techniques will lead to even higher accuracy in the future.







\subsection{Performance on Regular Cross-validation}
Table \ref{tab:comparison 1} presents the cross-validation results for PLD-Tree under the leave-one-protein-out approach. In this section, we assess unclassified cross-validation, where each protein, including mutant variants, is treated as an independent entity. Unlike our previous strategy of grouping proteins, the mutant dataset is incorporated directly without clustering, ensuring that each protein is evaluated uniquely during training. We tested PLD-Tree’s performance using 10-fold cross-validation on combined datasets (PDBbind V2020+SKEMPI v2 mt and PDBbind V2020+SKEMPI v2 wt\&mt), and all other results are provided in Supporting Information. Those datasets are improved $R_p$ from 0.78 to 0.87 and PDBbind V2020+SKEMPI v2 wt\&mt from 0.83 to 0.88 with MAEs of 1.049 kcal/mol and 1.037 kcal/mol. Augmenting the wild-type data in this independent scheme did not further improve results, likely because the mt dataset already captured sufficient variability from those proteins. These findings suggest that enlarging the dataset, whether by including mt or wt complexes, generally enhances performance, underscoring the benefits of incorporating diverse protein complexes into the model.

\subsection {Performance between Different Features}

Understanding how different feature types influence PPI binding affinity predictions is essential for guiding future feature engineering in protein design. To explore this, we examined how well our model’s predictions mimic the experimental data distributions when trained on distinct categories of features. Specifically, we compared the performance of our model using PPI structural/topological features, ESM sequence embeddings, and biophysical descriptors on the same dataset. Table \ref{tab:dission_subsection_1} summarizes these results, including performance metrics obtained through 10-fold cross-validation on the V2020+wt+mt dataset for each feature subset.

Our findings indicate that auxiliary
biophysical features generally outperform topological features when the dataset is smaller (e.g., V2020 alone). Additional Biophysical descriptors likely provide more direct, empirically grounded information about intermolecular interactions, making them particularly valuable when training data is limited. However, as we scale up to larger datasets (V2020+wt+mt), the relative advantage of biophysical features diminishes. Under these conditions, topological and ESM-based features can leverage the richer data environment more effectively. The ESM embeddings, trained on extensive protein sequence databases, provide a robust evolutionary and sequence-level context that complements the structural information captured by topological features.

In essence, each type of characteristic contributes unique information on the underlying determinants of protein-protein binding. Auxiliary features like biophysical features offer a direct link to physicochemical properties, ESM embeddings provide a broad sequence-level context, and topological representations encode essential structural details. By understanding how these feature categories shape model performance, we can make more informed decisions about integrating and prioritizing features in future predictive modeling efforts.

\begin{table}[h!]
    \centering
    \caption{Performance for $R_p$ and MAE by different features on 
 the considered datasets}
    \begin{tabular}{|c|c|c|c|c|c|c|c|c|}
        \hline
         & \multicolumn{2}{c|}{{V2020}} & \multicolumn{2}{c|}{V2020+wt} & \multicolumn{2}{c|}{{V2020+mt}} & \multicolumn{2}{c|}{{V2020+wt+mt}} \\ \hline


    \rowcolor{lightgray} 10fold & $R_p$ & MAE & $R_p$ & MAE & $R_p$ & MAE & $R_p$ & MAE \\ \hline
        {Auxiliary} & 0.6396& 1.609 & 0.6572 & 1.557 & 0.7149  & 1.659 & 0.7732 & 1.620 \\ \hline
        {PPI} &0.5898 & 1.707& 0.6589& 1.557 &0.7221  &1.694 &0.7996 & 1.515\\ \hline
        {ESM} & 0.6434 & 1.588 &  0.6814 & 1.458 &0.7699  & 1.522 &0.8059 & 1.419 \\ \hline
        {All} & 0.6718 & 1.560 &  0.7172& 1.402  & 0.7841 & 1.572    & 0.8276 &1.412 \\ \hline

    \end{tabular}
    
    \label{tab:dission_subsection_1}
\end{table}
It is noteworthy that, under 10-fold cross-validation, datasets containing single-mutation complexes consistently enhance the model’s predictive performance. This improvement can be attributed to the increased diversity introduced by various single mutations in the wild-type complexes, enabling the model to capture a broader range of variability and underlying patterns. In contrast, the other models discussed in the previous section primarily account for differences across distinct PDB structures rather than variations within the same PDB structure when specific testing data are selected. Furthermore, we observed that the v2020+wt+mt dataset significantly outperforms the V2020 dataset alone, as evidenced by a substantially lower MAE in the model’s predictions. This reduction in MAE highlights the enhanced performance achieved by incorporating single-mutation data, underscoring the value of diverse mutation information in improving predictive accuracy.

\subsection{Results based on Other Machine Learning Methods}
In addition to the GBDT model, we explored the use of neural network architectures, specifically Multilayer Perceptrons (MLPs), as an alternative approach for predicting PPI binding affinities. Despite the flexibility and theoretical advantages of neural networks in capturing complex nonlinear relationships, our initial experiments revealed that the MLP did not surpass the performance of the GBDT model. This outcome may be attributed to several factors, including the relatively limited size of our training dataset and the specific architectural choices of the neural network, which may not have been optimal for this particular application. Nevertheless, these findings highlight the potential of neural network-based models for future studies. With advancements in model architecture design and the availability of larger, more diverse datasets, neural networks could achieve superior performance by better leveraging the intricate patterns and high-dimensional feature spaces inherent in PPI data. Future research should focus on experimenting with more sophisticated deep learning architectures, such as convolutional neural networks (CNNs) or graph neural networks (GNNs), and on expanding the dataset size to fully realize the capabilities of neural network models in enhancing PPI binding affinity predictions.

\section{Conclusion}

In this work, we introduced PLD-Tree, a binding free energy predictor specifically tailored for protein–protein complexes. Through comprehensive evaluations on three widely recognized datasets—including SKEMPI v2 and, for the first time, PDBbind v2020 used as a training set for PPI binding affinity predictions—PLD-Tree demonstrated consistent and robust performance. Compared to existing empirical predictors, our approach not only achieved superior predictive accuracy but also maintained remarkable stability across multiple benchmark scenarios.

Additionally, we examined PLD-Tree’s capabilities on various test sets, further highlighting its reliability and adaptability. To facilitate broader application, we have made an open-source version of PLD-Tree available on GitHub, featuring functionalities for screening protein complexes and engineering amino acid compositions at the interface to enhance binding affinity or identify critical mutants that may compromise stability. This versatility positions PLD-Tree as a valuable tool in the design of novel biologics and therapeutic compounds, underscoring its potential impact in advancing protein–protein interaction research and pharmaceutical development.


\section{Methods}
In this section, we provide an overview of algebraic topology \cite{hatcher2002algebraic,massey2019basic,edelsbrunner2010computational,zomorodian2004computing} that are frequently employed to capture the geometry and topology of molecular structures. By applying these complexes to PPI binding interfaces, we can construct multi-scale representations that serve as the foundation for both persistent homology and the persistent Laplacian.

\subsection{Alpha Shapes}
Alpha shapes extend the concept of a convex hull to reveal finer geometric details of a point cloud, making them particularly suitable for representing molecular surfaces and interfaces. Formally, given a set of points $S \subset \mathbb{R}^n$, one first constructs the Delaunay triangulation of $ S $. The alpha shape of $ S $ at scale parameter $\alpha$ is then derived by selectively removing simplices whose circumscribing spheres exceed radius $\alpha$. In other words, as $\alpha$ varies from small to large values:
\begin{itemize}
\item For very small $\alpha$, the alpha shape captures only the tightest clusters of points, isolating them into small connected components.
\item As $\alpha$ increases, more simplices remain in the alpha shape, gradually adding bridges, faces, and volumes that reflect the evolving structure of the underlying point set.
\item Eventually, for a sufficiently large $\alpha$, the alpha shape includes the entire Delaunay triangulation, effectively replicating the convex hull of $S$.
\end{itemize}

This progression from coarse to fine representations provides a filtration—a nested sequence of complexes—on which one can compute topological invariants (e.g., connected components, loops, and voids). In the context of PPI, alpha shapes are valuable for identifying cavities, tunnels, and other structural features that may be crucial to binding. By examining how these features persist or disappear over changes in $\alpha$, one can obtain insights into the stability and geometry of protein interfaces.

\subsection{Rips Complex}
In topology, the Rips complex (also called the Vietoris–Rips complex) is another fundamental construction for translating distance information in a point set into a simplicial complex. Unlike alpha shapes, which rely on Delaunay triangulation, the Rips complex directly uses pairwise distances among points in a metric space.

Formally, let  $S \subset M$  be a set of points in a metric space $(M, d)$. For a chosen distance threshold $\epsilon > 0$, the Rips complex  $R_{\epsilon}(S)$  is defined as follows:
\begin{itemize}
\item Each point in $ S $ corresponds to a vertex in the complex.
\item A (k-dimensional) simplex is formed by any $k+1$ points $(v_0, v_1, \dots, v_k) \subset S$ if and only if the distance  $d(v_i, v_j) \le \epsilon  \text{ for all } 0 \le i < j \le k$.
\end{itemize}

By increasing $\epsilon$ from small to large, one obtains a filtration of Rips complexes. Initially, when $\epsilon$ is very small, each point stands alone with no edges. As $\epsilon$ grows, edges, triangles, and higher-dimensional simplices appear, linking points into connected components and revealing loops, cavities, and other topological features. This scale-dependent perspective is integral to persistent homology, as it tracks how topological features emerge and vanish over the course of filtration. In molecular modeling, the Rips complex offers a direct way to capture multi-scale geometry from pairwise atomic or residue distances, making it a powerful tool for analyzing the binding interfaces of protein complexes.

\subsection{Simplicial Complex and Filtration}
An abstract simplicial complex is a finite collection of sets of points (that is, atoms) $K =\{\sigma_i\}_i$, where the elements in $\sigma_i$ are called vertices and $\sigma_i$ is called a k-simplex if it has $k+1$ distinct vertices. If $\tau \subseteq \sigma_i$ for $\sigma_i \in K$ indicates that $\tau \in K$, and that the non-empty intersection of any two simplices $\sigma_1,\sigma_2 \in K$, is a face of both $\sigma_1$ and $\sigma_2$.

In practice, it is favorable to characterize point clouds or atomic positions in various spatial scales rather than in a fixed scaled simplicial complex representation. To construct a scale-changing simplicial complex, consider a function $f: K \to \mathbb{R}$ that satisfies $f(\tau)\le f(\sigma)$ whenever $\tau \subseteq \sigma$. 
Given a real value, $x,f$ induces a subcomplex of $K$ by constructing a sub-level set, $K(x)=\{\sigma \in K|f(\sigma) \le x\}$. As $K$ is finite, the range of $f$ is also finite and the induced subcomplexes, when ordered, form a filtration of $K$,
\[\emptyset \subset K(x_1)\subset K(x_2) \subset ... \subset K(x_l) = K\]

There are many constructions of $K$ and one that is widely used for point clouds is the Rips complex. Given $K$ as the collection of all possible simplices from a set of atomic coordinates until a fixed dimension, the filtration function 
is defined as $f_{\text{Rips}}(\sigma) = \max\{d(v_i,v_j)|v_i,v_j \in \sigma\}$ for $\sigma \in K$, where $d$ is a predefined distance function between the vertices; for example, $D_e$. In practice, an upper bound of the filtration value is set to avoid an excessively large simplicial complex. Another efficient construction called the alpha complex is often used to characterize geometry, and we denote the filtration function by $f_{\alpha}:DT(X) \to \mathbb{R}$, where $DT(X)$ is the simplicial complex that is induced by the Delaunay triangulation of the set of atomic coordinates, $X$. The filtration function is defined as $f_{\alpha}(\sigma) = \max\{\frac{1}{2}D_e(v_i,v_j)|v_i,v_j \in \sigma\}$ for $\sigma \in DT(X)$. Back to molecular structures, the filtration of simplicial complexes describes the topological characteristics of interaction hypergraphs under various interaction range assumptions.

\subsection{Homology Groups}  In the context of singular homology, a homology group of a simplicial complex topologically represents hole-like structures of various dimensions. Given a simplicial complex \( K \), a \( k \)-chain is a finite formal sum of \( k \)-simplices in \( K \); that is, \( \sum_i a_i\sigma_i \). For simplicity, we choose coefficients \( a_i \) from \( \mathbb{Z}_2 \). The \( k \)-th chain group, denoted \( C_k(K) \), comprises all \( k \)-chains with addition induced by the addition of coefficients.

A boundary operator \( \partial_k: C_k(K) \rightarrow C_{k-1}(K) \) connects chain groups of different dimensions by mapping a chain to the alternating sum of its codimension-1 faces. For simplices, this operator is defined as follows:
\[
\partial_k(\{v_0, \ldots, v_k\}) = \sum_{i=0}^k (-1)^i [v_0, \ldots, \hat{v_i}, \ldots, v_k],
\]
where \( \hat{v_i} \) signifies that vertex \( v_i \) is omitted.
The \( k \)-th cycle group, denoted \( Z_k(K) \), is the kernel of \( \partial_k \), and its elements are called \( k \)-cycles. The \( k \)-th boundary group, \( B_k(K) \), is the image of \( \partial_{k+1} \). By the property \( \partial_k \circ \partial_{k+1} = 0 \), \( B_k(K) \) is a subgroup of \( Z_k(K) \).
The \( k \)-th homology group, \( H_k(K) \), is defined as the quotient group \( Z_k(K) / B_k(K) \). The equivalence classes in \( H_k(K) \) correspond to \( k \)-dimensional holes in \( K \) that cannot be deformed into each other by the boundary of a subcomplex.

Given a filtration as mentioned earlier, in addition to characterizing the homology group at each step \( H_k(K(x_i)) \), it is important to track the persistence of topological features throughout the sequence. Viewing \( H_k(K(x_i)) \) as vector spaces with inclusion map-induced linear transformations gives a persistence module:
\[
H_k(K(x_1)) \rightarrow H_k(K(x_2)) \rightarrow \ldots \rightarrow H_k(K(x_\ell)).
\]
An interval module with respect to \([b, d)\), denoted \( I_{[b,d)} \), is defined as a collection of vector spaces \( \{V_i\} \) connected by linear maps \( f_i: V_i \rightarrow V_{i+1} \). Here, \( V_i = \mathbb{Z}_2 \) for \( i \in [b, d) \) and \( V_i = 0 \) otherwise. The map \( f_i \) is the identity map when possible and zero otherwise.

The persistence module can be decomposed as a direct sum of interval modules \( \bigoplus_{[b,d) \in B} I_{[b,d)} \). Each \( I_{[b,d)} \) corresponds to a homology class that appears at filtration value \( b \) (birth) and disappears at filtration value \( d \) (death). The collection of these pairs, \( B \), records the evolution of \( k \)-dimensional holes with varying filtration parameters, and thus captures the topological configuration of the input point cloud under different interaction ranges, especially when using a distance-based filtration.

\section{Data and Code Availability}
All data and code utilized in this study will be made publicly accessible on GitHub upon the acceptance of this paper. The GitHub repository will include:

\begin{itemize}
\item Datasets: Processed versions of the PDBbind v2020 and SKEMPI v2 datasets, along with any additional datasets used in our analysis.
\item Code: Scripts for data preprocessing, feature extraction, model training, and evaluation of the PLD-Tree model.
\item Documentation: Detailed instructions and guidelines to facilitate replication of our results, including dependencies and usage examples.
\item Trained Models: Pre-trained PLD-Tree models for immediate use and benchmarking.
\end{itemize}

\section{Supporting Information}
Some experiments and performances utilized in this work will be provided in the supporting information. And it will include.
\begin{itemize}
    \item The tables for $R_p$ and MAE between experimental and predicted binding affinities on S630, S55, and S685
    \item The table and figure for $R_p$ and MAE for different datasets V2020+mt, V2020+wt+mt and feature sets 
    \item The figures for $R_p$ and MAE for S177
    \item The tables for different hyper-parameters for testing

\section*{Acknowledgments}
 J. C. was partially supported by Arkansas biosciences institute seed grant.  C. W. was partially supported   by the National Science Foundation under award DMS-2206332. 
    
\end{itemize}
\bibliographystyle{plain}
\bibliography{main}
\end{document}